\RequirePackage{fixltx2e}
\documentclass[aps,prb,reprint,floatfix]{revtex4-1}
\usepackage{amsmath,amsfonts,amssymb} %
\usepackage{mathtools} %
\usepackage{wasysym} %
\usepackage{bm} %
\usepackage[bbgreekl]{mathbbol} %
\usepackage{graphicx} %
\usepackage[dvipsnames,table]{xcolor} %
\usepackage{tabularx} %
\usepackage[acronym]{glossaries} %
\usepackage{braket} %

\usepackage{fancyhdr} %
\usepackage{microtype} %
\usepackage{natmove}

\usepackage[byname]{smartref} %
\usepackage[breaklinks, %
colorlinks, linkcolor=Blue, citecolor=Blue, urlcolor=Blue]{hyperref} %
\usepackage[version=3]{mhchem}
\AtBeginDocument{\usepackage{booktabs}} 

\newcommand{\BC}{\color{black} } %
\newcommand{\eq}[1]{Eq.~(\ref{#1})} %
\newcommand{\fig}[1]{Fig.~\ref{#1}} %

\def\be{\begin{equation}} %
\def\ee{\end{equation}} %
\def\bea{\begin{eqnarray}} %
\def\eea{\end{eqnarray}} %

\newacronym{QPE}{QPE}{quantum phase estimation} %
\newacronym{VQE}{VQE}{variational quantum eigensolver} %
\newacronym{UCC}{UCC}{unitary coupled cluster} %
\newacronym{QCC}{QCC}{qubit coupled cluster} %
\newacronym{FCI}{FCI}{full configurational interaction} %
\newacronym{CASCI}{CASCI}{complete active space configurational
  interaction} %
\newacronym{JW}{JW}{Jordan--Wigner} %
\newacronym{BK}{BK}{Bravyi--Kitaev} %
\newacronym[longplural={degrees of freedom}, %
firstplural={degrees of freedom (DOF)}, plural={DOF}]{DOF}{DOF}{degree
  of freedom} %
\newacronym[longplural={equations of motion}, %
firstplural={equations of motion (EOM)}, %
plural={EOM}]{EOM}{EOM}{equation of motion} %
\newacronym{PES}{PES}{potential energy surface} %
\newacronym{CI}{CI}{configuration interaction} %
\newacronym{QMF}{QMF}{qubit mean-field} %
\newacronym{SQP}{SQP}{sequential quadratic programming} %
\newacronym{RHF}{RHF}{restricted Hartree--Fock}

\begin{document}

\author{Artur F. Izmaylov} 
\email{artur.izmaylov@utoronto.ca}
\affiliation{Department of Physical and Environmental Sciences,
  University of Toronto Scarborough, Toronto, Ontario, M1C 1A4,
  Canada; and Chemical Physics Theory Group, Department of Chemistry,
  University of Toronto, Toronto, Ontario, M5S 3H6, Canada}

\author{Tzu-Ching Yen} 
\affiliation{Department of Physical and Environmental Sciences,
  University of Toronto Scarborough, Toronto, Ontario, M1C 1A4,
  Canada; and Chemical Physics Theory Group, Department of Chemistry,
  University of Toronto, Toronto, Ontario, M5S 3H6, Canada}

\author{Ilya G. Ryabinkin} 
\affiliation{OTI Lumionics Inc., 100 College Street 351, Toronto,
  Ontario\, M5G 1L5, Canada} %
  
\title{Revising measurement process in the variational quantum eigensolver: Is it possible 
to reduce the number of separately measured operators?}

 %


\begin{abstract}
Current implementations of the Variational Quantum Eigensolver (VQE) 
technique for solving the electronic structure problem involve
splitting the system qubit Hamiltonian into parts whose elements commute within their single qubit 
subspaces. The number of such parts rapidly grows with the size of the molecule. This 
increases the computational cost and can increase uncertainty in the measurement of 
the energy expectation value because elements from different parts need to be measured independently. 
To address this problem we introduce a more efficient partitioning of 
the qubit Hamiltonian using fewer parts that need to be measured separately. 
The new partitioning scheme is based on 
two ideas: 1) grouping terms into parts whose eigenstates have a single-qubit product structure, 
and 2) devising multi-qubit unitary transformations for the Hamiltonian or its parts to produce less entangled 
operators. The first condition allows the new parts to be measured in the number of 
involved qubit consequential one-particle measurements. Advantages of the new partitioning scheme 
resulting in severalfold reduction of separately measured terms are illustrated on the H$_2$ and LiH problems.  
\end{abstract}

\glsresetall

\maketitle

\section{Introduction}

One of the most practical schemes for solving the electronic structure problem 
on current and near-future universal quantum computers is the \gls{VQE}
method.\cite{Peruzzo:2014/ncomm/4213, Jarrod:2016/njp/023023, Wecker:2015/pra/042303,Olson:2017ud,McArdle:2018we} 
This approach involves the following steps:
1) reformulating the electronic Hamiltonian ($\hat H_e$) in the second quantized form,
2) transforming $\hat H_e$ to the qubit form ($\hat H_q$) 
by applying iso-spectral fermion-spin transformations such as  
\gls{JW}~\cite{Jordan:1928/zphys/631, AspuruGuzik:2005/sci/1704} or
more resource-efficient \gls{BK}~\cite{Bravyi:2002/aph/210, Seeley:2012/jcp/224109,
  Tranter:2015/ijqc/1431, Setia:2017/ArXiv/1712.00446,
  Havlicek:2017/pra/032332}, 
  3) solving the eigenvalue problem for $\hat H_q$ by variational optimization of
unitary transformations for a qubit wavefunction. 
The last step uses hybrid quantum-classical technique where
a classical computer suggests a trial unitary transformation $U$, and its 
quantum counterpart provides energy 
expectation value of $E_U = \bra{\Psi_0}U^\dagger\hat H_q U\ket{\Psi_0}$,
here $\ket{\Psi_0}$ is an initial qubit wavefunction (it is frequently taken as an uncorrelated 
product of all spin-up states of individual qubits). The two steps, on 
classical and quantum computers, are iterated till convergence. The VQE 
was successfully implemented on several quantum computers and 
used for few small molecules up to BeH$_2$.\cite{Kandala:2017/nature/242} 
 
One of the big problems of the \gls{VQE} is that to calculate $E_U$, the quantum 
computer measures parts of $H_q$ rather than the whole $H_q$ on the $U\ket{\Psi_0}$ wavefunction.
This stems from technological restrictions of what can be currently measured on available architectures.
Dramatic consequences of this restriction can be easily understood on the following simple example. 
Let us assume that $\hat H_q = \hat A+\hat B$, where $\hat A$ and $\hat B$ are measurable 
components of $\hat H_q$ and $[\hat A,\hat B] \ne 0$, otherwise they could be measured at 
the same time at least in principle. The actual hardware restrictions on measurable components
are somewhat different and will be discussed later, for this illustration these differences are not important.
Even if one has an exact eigenstate of  $\hat H_q$, $U\ket{\Psi_0}$, measuring it on $\hat A$ 
or $\hat B$ would not give a certain 
result because $\hat A$ and $\hat B$ do not commute with $\hat H_q$. Thus, one would not be able to 
distinguish the exact eigenstate from other states by its zero variance from the energy expectation value. 
The origin of the discrepancy between quantum uncertainty given by variance ($Var$) 
of $\hat H_q$ (true uncertainty) and by sum of variances for $\hat A$ and $\hat B$ 
is neglect of the covariance part ($Cov$)
\bea\notag
Var(\hat H_q) &=& Var(\hat A) + Var(\hat B) \\ \label{eq:Hvar}
&&+ Cov(\hat A,\hat B) + Cov(\hat B,\hat A), \\
Var(\hat A) &=& \langle\hat A^2\rangle - \langle\hat A\rangle^2, \\ \label{eq:cov}
Cov(\hat A,\hat B) &=& \langle\hat A \hat B\rangle - \langle\hat A\rangle\langle\hat B\rangle.
\eea 
Thus, even though the $\hat H_q$ average is equal to averages of $\hat A$ and $\hat B$,
the true quantum uncertainty of $\hat H_q$ is overestimated by a sum of variances for $\hat A$ and $\hat B$. 
Moreover, the number of measurements to sample $\hat A$ and $\hat B$ is twice as many as that for 
$\hat H_q$ if the eigenstate nature of $U\ket{\Psi_0}$ is not known {\it a priori}.

{\BC Variance of any Hamiltonian depends only on the Hamiltonian and the wavefunction, but if one 
approximates the variance using only variances of Hamiltonian parts and neglects covariances between the parts 
the result of such an approximation will depend on the partitioning. Importantly, the sum of 
variances for the Hamiltonian parts can either under- or overestimate the true Hamiltonian variance.  
To see how ignoring covariances can erroneously make estimates of the uncertainty arbitrarily small 
consider an artificial example, where the Hamiltonian variance  
is measured as $n$ independent measurements of its $\hat H_q/n$ identical parts. 
Due to the linear scaling of the variance sum with $n$ and the inverse quadratic scaling  
of variances of individual terms with $n$, the overall scaling of the variance is inversely proportional to $n$ 
and can be made arbitrarily small by choosing large $n$. This follows from a wrong assumption
 that parts ($\hat H_q/n$) are independent and covariances between them are zero.}
    
Generally, the number of non-commuting terms in $\hat H_q$ grows with the size of the original 
molecular problem, and the total uncertainty from the measurement of individual terms will increase. 
This increase raises the standard deviation of the total measurement process and leads to  
a large number of measurements to reach convergence in the energy expectation value.  
The question we would like to address is whether it is possible to reduce the number of 
the $\hat H_q$ terms that needs to be measured separately?     
 
In this paper we introduce a new systematic approach to decreasing uncertainty
of the expectation energy measurement. We substitute the conventional measurement 
partitioning of the Hamiltonian 
to groups of qubit-wise commuting operators\cite{Kandala:2017/nature/242,Ryabinkin:2018/cvqe} 
by partitioning to terms whose eigenstates 
can be found exactly using the mean-field procedure. Due to more general structure of 
such terms the Hamiltonian can be split into a fewer number of them. 
Interestingly, the general operator conditions on such mean-field terms have not been found in the 
literature and have been derived in this work for the first time.
To decrease the number of these terms even further, we augment the mean-field treatment 
with few-qubit unitary transformations that allows us to measure few-qubit entangled terms. 
Measurement of newly introduced terms requires the scheme appearing in the cluster-state
quantum computing,\cite{Nielsen:2006ve,Mantri:2017el} 
it is qubit-wise measurement with use of previous measurement results to define what single-qubit 
operators to measure next.   

\section{Theory}
\label{sec:theory}

\subsection{Qubit Hamiltonian}

 In order to formulate the electronic structure problem for a
quantum computer that operates with qubits (two-level systems), the
electronic Hamiltonian needs to be transformed iso-spectrally to its
qubit form. This is done in two steps. First, the
second quantized form of $\hat H_e$ is obtained
\begin{equation}
  \label{eq:qe_ham}
  \hat H_e = \sum_{pq} h_{pq} {\hat a}^\dagger_p {\hat a}_q + \frac{1}{2}\sum_{pqrs}
  g_{pqrs} {\hat a}^\dagger_p {\hat a}^\dagger_q {\hat a}_s {\hat a}_r,
\end{equation}
where ${\hat a}_p^\dagger$ (${\hat a}_p$) are fermionic creation
(annihilation) operators, $h_{pq}$ and $g_{pqrs}$ are one-
and two-electron integrals in a spin-orbital
  basis.\cite{Helgaker:2000} This step has polynomial complexity and
is carried out on a classical computer. Then, using the
\gls{JW}~\cite{Jordan:1928/zphys/631, AspuruGuzik:2005/sci/1704} or
more resource-efficient \gls{BK}
transformation~\cite{Bravyi:2002/aph/210, Seeley:2012/jcp/224109,
  Tranter:2015/ijqc/1431, Setia:2017/ArXiv/1712.00446,
  Havlicek:2017/pra/032332}, the electronic Hamiltonian is converted
iso-spectrally to a qubit form
\begin{equation}
  \label{eq:spin_ham}
  \hat H_q = \sum_I C_I\,\hat P_I,
\end{equation}
where $C_I$ are numerical coefficients, and $\hat P_I$ are
Pauli  ``words", products of Pauli operators of different qubits 
\begin{equation}
  \label{eq:Pi}
  \hat P_I = \cdots \hat \sigma_{2}^{(I)} \, \hat \sigma_{1}^{(I)},
\end{equation}
$\hat \sigma_i^{(I)}$ is one of the $\hat x,\hat y,\hat z$ Pauli
operators for the $i^{\rm th}$ qubit. The number of qubits $N$ is equal 
to the number of spin-orbitals used in the second quantized form [\eq{eq:qe_ham}].
Since every fermionic operator is substituted by a product of Pauli operators in 
both JW and BK transformations, the total number of Pauli words in $\hat H_q$ 
scales as $N^4$. 

\subsection{Conventional measurement} 

In the conventional VQE scheme the $\hat H_q$ is separated into sums of 
qubit-wise commuting (QWC) terms, 
\bea\label{eq:split}
\hat H_q &=&  \sum_{n} \hat A_n,~ [\hat A_n,\hat A_k]_{\rm qw} \ne 0,~{\rm if}~n\ne k \\
 \hat A_n &=& \sum_{I} C_{I}^{(n)}\,\hat P_{I}^{(n)}, ~ [\hat P_{I}^{(n)},\hat P_{J}^{(n)}]_{\rm qw} = 0, ~\forall I \& J.
\eea
Here $[\hat P_{I}^{(n)},\hat P_{J}^{(n)}]_{\rm qw}$ denotes a qubit-wise commutator of two Pauli words, 
it is zero only if all one-qubit operators in $\hat P_{I}^{(n)}$ commute with their counterparts in $\hat P_{J}^{(n)}$.
Clearly, if $[\hat P_{I}^{(n)},\hat P_{J}^{(n)}]_{\rm qw}=0$ then the normal commutator is 
$[\hat P_{I}^{(n)},\hat P_{J}^{(n)}]=0$. The 
opposite is not true, a simple example is $[\hat x_1\hat x_2, \hat y_1\hat y_2] = 0$ but 
$[\hat x_1\hat x_2, \hat y_1\hat y_2]_{\rm qw} \ne 0$. We will not be using non-zero results of 
the qubit-wise commutator and therefore their exact values are not important, but it is assumed that 
$[.,.]_{\rm qw}$ is bi-linear for both operators.

Partitioning of the $H_q$ in Eq.~(\ref{eq:split}) allows one to measure all Pauli words within each $\hat A_n$ term 
in a single set of $N$ one-qubit measurements. For every qubit, it is known from the form of
$\hat A_n$, what Pauli operator needs to be measured. The advantage of this scheme is that 
it requires only single-qubit measurements, which are technically easier than multi-qubit measurements.
The disadvantage of this scheme is that the Hamiltonian may require to measure too many $\hat A_n$ 
terms separately.   

A natural extension of partitioning in Eq.~(\ref{eq:split}) is to sum more general terms 
\bea\label{eq:splitMF}
\hat H_q &=& \sum_n \hat H_n^{\rm (MF)},
\eea
with the condition that $\hat H_n^{\rm (MF)}$ eigenstates can be presented in a single product 
form of single-qubit wavefunctions. In other words, the eigenstates of 
the $\hat H_n^{\rm (MF)}$  fragments are unentangled and can be obtained using a 
mean-field procedure. This condition would allow measurement of each $\hat H_n^{\rm (MF)}$
fragment qubit after a qubit. However, to perform the new splitting we need an exact definition 
of the mean-field (MF) Hamiltonian so that we can recognize these new blocks within the total Hamiltonian.

\subsection{Mean-field Hamiltonians}

What is the most general form of a qubit Hamiltonian whose eigenstates can be presented as  
single factorized products of one-qubit wavefunctions? 
Note that, the well-known example of such Hamiltonians, separable operators
\bea\label{eq:sep}
\hat H_S(1,...,N) = \sum_{i=1}^{N} \hat h_i(i) 
\eea
is a particular class that does not provide the most general form. In other words, there are many more 
Hamiltonians that are not separable but are still in the MF class, one simple example is 
\bea\label{eq:mf}
\hat H_{\rm MF} (1,2) = \hat x_2 + \hat z_1 \hat y_2, 
\eea
which does not follow the form of \eq{eq:sep} but whose eigenstates,
$\ket{+_z}_1\ket{\pm_{x+y}}_2$ and $\ket{-_z}_1\ket{\pm_{x-y}}_2$,\footnote{Here, we use the notation 
$\ket{\pm_\sigma}_n$ for the $n^{th}$ qubit eigenstates of $\sigma$ one-particle operator with  
$\pm 1$ eigenvalues.} 
are unentangled products.  

We formulate the general criterion for a Hamiltonian $H(1,...N)$ to be in the MF class  
as follows. There should exist $N$ one-particle 
operators $\{\hat O_k(k)\}_{k=1}^{N}$\footnote{To simplify the notation 
we use freedom in qubit enumeration and assume that we work with 
the qubit enumeration that follows the described reductive sequence.} 
that commute $[\hat O_k,\hat H_{N-k+1}]=0$
with the system of $N$ Hamiltonians $\{\hat H_{N-k+1}\}_{k=1}^N$ 
constructed in the following way that we will refer as a {\it reductive chain}: 
\bea\notag
&1:&~ \hat H_N = \hat H, \\ \notag
&2:&~ \hat H_{N-1} = \bra{\phi_1} \hat H_N \ket{\phi_1},\\\notag
&&... \\\label{eq:chain}
&N:&~ \hat H_1 = \bra{\phi_{N-1}} \hat H_{2} \ket{\phi_{N-1}},
\eea   
where $\hat O_{k}\ket{\phi_{k}} = \lambda_{k} \ket{\phi_{k}}$. The final operator in this chain is a 
one-particle operator that commutes with itself and defines $\hat O_N= \hat H_1$. 
The proof of this criterion can be found in appendix A. 
It is easy to see that $\ket{\Psi} = \prod_{k=1}^{N} \ket{\phi_k}$ is an eigenfunction of 
$\hat H$. Clearly, separable Hamiltonians are in the MF class 
because for them, $\hat O_k$'s can be taken as $\hat h_k(k)$ from \eq{eq:sep}. 
However, note that because the system of $\hat O_k$ operators is required to commute 
not with $\hat H$ but with the reduced set of Hamiltonians, the formulated criterion goes beyond 
separable Hamiltonians. 

A general procedure to determine whether a particular qubit Hamiltonian 
$\hat H$ is in the MF class or not requires finding all $N$ one-particle operators $\hat O_k$.      
The procedure starts with a check whether there is at least one qubit $k$ for which 
\bea\label{eq:comm}
[\hat H,(a \hat x_k+b \hat y_k + c\hat z_k)] = 0 
\eea 
can be achieved by choosing non-zero vector $(a,b,c)$. Once the first operator 
$\hat O_1(k) = a \hat x_k+b \hat y_k + c\hat z_k$ is found its eigenstates can be integrated out 
to generate $\hat H_{N-1}$, and the procedure can be repeated to find $\hat O_2$
that commutes with $\hat H_{N-1}$.

\subsection{Measurement of mean-field Hamiltonians} 

Measuring an $N$-qubit mean-field Hamiltonian can be done by performing a single set of sequential 
$N$ one-qubit measurements. Each qubit projective measurement in this set will collapse the 
measured wavefunction to an eigenstate of the corresponding single qubit operator. The single qubit 
operators that need to be measured are $\hat O_k$'s operators. The definition of one particle operators 
may depend on the result of the previous measurement. Let us consider the mean-field 
Hamiltonian in \eq{eq:mf}: $\hat O_1(1) = \hat z_1$, and $\hat O_2(2) = \hat x_2\pm\hat y_2$, where 
$\pm$ is determined by the eigenfunction chosen from the $\hat O_1$ spectrum to generate 
the $\hat H_1 = \bra{\phi_1^{\pm}}H_{\rm MF}\ket{\phi_1^{\pm}}$ in the chain of \eq{eq:chain}. 
This ambiguity does not allow one to present $\hat H_{\rm MF}$ as an operator with all 
qubit-wise commuting components. An attempt of this can be done by inserting 
the projectors on the eigenstates of $\hat z_1$ instead of the operator:
\bea
\hat H_{\rm MF} &=& (\hat x_2 + \hat y_2)\ket{\phi_1^{+}}\bra{\phi_1^{+}} + 
(\hat x_2 - \hat y_2) \ket{\phi_1^{-}}\bra{\phi_1^{-}} \\
&=& [(\hat x_2 + \hat y_2)(1+ \hat z_1) + (\hat x_2 - \hat y_2)(1-\hat z_1)]/2,
\eea    
where $\hat z_1 \ket{\phi_1^{\pm}} = \pm \ket{\phi_1^{\pm}}$, and even though the projectors 
onto the $\ket{\phi_1^{\pm}}$ eigenstates commute, the $(\hat x_2 \pm \hat y_2)$ parts do not.     

\begin{figure}[h!]
  \centering %
  \includegraphics[width=0.85\columnwidth]{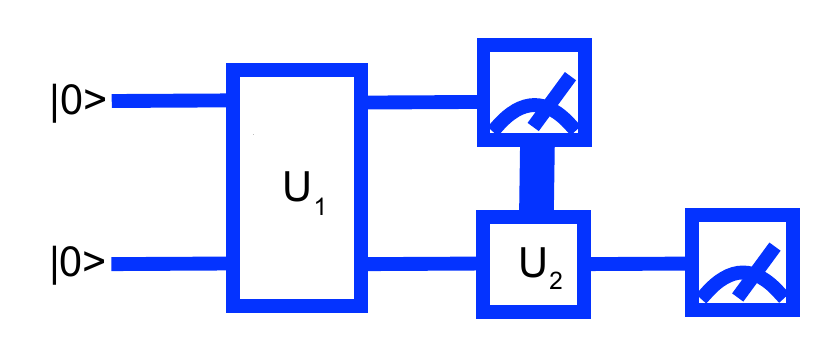}
  \caption{Measurement where the second qubit is rotated by $U_2$ 
  depending on the result of the first qubit measurement.}
  \label{fig:MFm}
\end{figure}

Therefore, the scheme for measuring the $\hat H_{\rm MF}$ will be as shown in Fig.~\ref{fig:MFm}.  
Note, that no matter how entangled the initial wavefunction is, measuring $\hat H_{\rm MF}$
does not require measuring $\hat x_2$  and $\hat z_1 \hat y_2$ separately as was done in 
the regular VQE scheme. 

{\BC
In practice, qubit-wise measurement using previous measurement results to define what single-qubit 
operators to measure next, or {\it feedforward} measurements, have been implemented in quantum computers
based on superconductor and photonics qubit architectures.\cite{nArriagada:2018ju,Prevedel:2007ca} 
The essential feasibility condition for  
 the feedforward measurement is that the delay introduced by 
measurements is much shorter than the qubit coherence time. For superconducting (photonics) qubit 
architectures this condition has been achieved with typical timescales for a measurement and 
coherence as 2 $\mu s$\cite{Riste:2015te} (150 ns\cite{Prevedel:2007ca}) and 
40 $\mu s$\cite{PhysRevLett.111.080502} (100 ms\cite{Korber:2017cu}), 
respectively.

\subsection{Mean-field partitioning}
\label{sec:parts}

Even though regular molecular qubit Hamiltonians are not guaranteed to be in the MF class,
it is always possible to split any $N$-qubit Hamiltonian into a sum of MF Hamiltonians. 
To see this, we will present a heuristic partitioning scheme that guarantees the MF partitioning. 

Our scheme uses ranking of all qubits $k=1,...,N$ based on a geometrical characteristic $l(k)$, 
which is defined as follows. For an arbitrary qubit $k$, the total Hamiltonian can be written as
\bea\label{eq:HO}
\hat H &=& \hat h_x \hat x_k + \hat h_y \hat y_k + \hat h_z \hat z_k+ \hat h_e 
\eea
where $\hat h_{x,y,z,e}$ are the residual operators that do not contain Pauli 
matrices for the $k^{\rm th}$ qubit. Assembling coefficients of Pauli words 
in operators $\hat h_{x,y,z}$ into vectors, $\bar{h}_{x,y,z}$, 
we build matrix $A_k = [\bar{h}_{x} \bar{h}_{y} \bar{h}_{z}]$ with dimensions $M$ by $3$, 
where $M$ is the number of different Pauli words in  $\hat h_{x,y,z}$ operators.  
To define $l(k)$, we evaluate matrix $S_k= A_k^{\dagger}A_k$ and assign 
$l(k) = \dim(\ker(S_k))$. Evaluating $S_k$ is equivalent to obtaining the 
overlap between three vectors $\bar{h}_{x,y,z}$ assuming the orthogonal basis, while  
the dimensionality of its kernel is the number of its zero eigenvalues. 

$l(k)$ allows one to answer a question on whether there is a transformation involving only the $k^{\rm th}$ qubit 
that can present $\hat H$ in one of two forms:
\bea \label{eq:O1}
\hat H &=& \hat h \hat O_k + \hat h_e, \\ \label{eq:O2}
\hat H &=& \hat h' \hat O'_k + \hat h'' \hat O''_k + \hat h_e,
\eea
where $\hat O_k,\hat O'_k,\hat O''_k$ are operators containing only the $k^{\rm th}$ qubit, and 
$\hat h,\hat h',\hat h''$ are the complementary operators that exclude the $k^{\rm th}$ qubit. 
The positive answers in the forms of \eq{eq:O1} and \eq{eq:O2} correspond to $l(k)=2$ and $l(k)=1$, respectively.  
$l(k)=2$ is equivalent to the MF condition of \eq{eq:comm}, with $\hat O_k = a \hat x_k+b \hat y_k + c\hat z_k$.
For $l(k)=1$, the MF treatment of the 
$k^{\rm th}$ qubit is not possible but using \eq{eq:O2} the $k^{\rm th}$ qubit dependence
in the Hamiltonian can be somewhat compactified. 
Coefficients for $\hat O_k,\hat O'_k,\hat O''_k$ and $\hat h,\hat h',\hat h''$
operators can be found from non-zero eigenvectors of $S_k$ (this process is detailed in appendix B).
 The negative answer to the question leaves 
$\hat H$ in the original form of \eq{eq:HO} and is equivalent to $l(k)=0$. 

The question about possible compactification of the $k^{\rm th}$ qubit dependence in the Hamiltonian 
has a simple geometric interpretation in terms of arrangement of three vectors $\bar{h}_{x,y,z}$. 
These multi-dimensional vectors can be linearly independent (\eq{eq:HO}), 
located within some plane (\eq{eq:O2}), or collinear to each other (\eq{eq:O1}), 
Fig.~\ref{fig:hvecs} illustrates all three cases. 
  \begin{figure}[h!]
  \centering %
  \includegraphics[width=0.75\columnwidth]{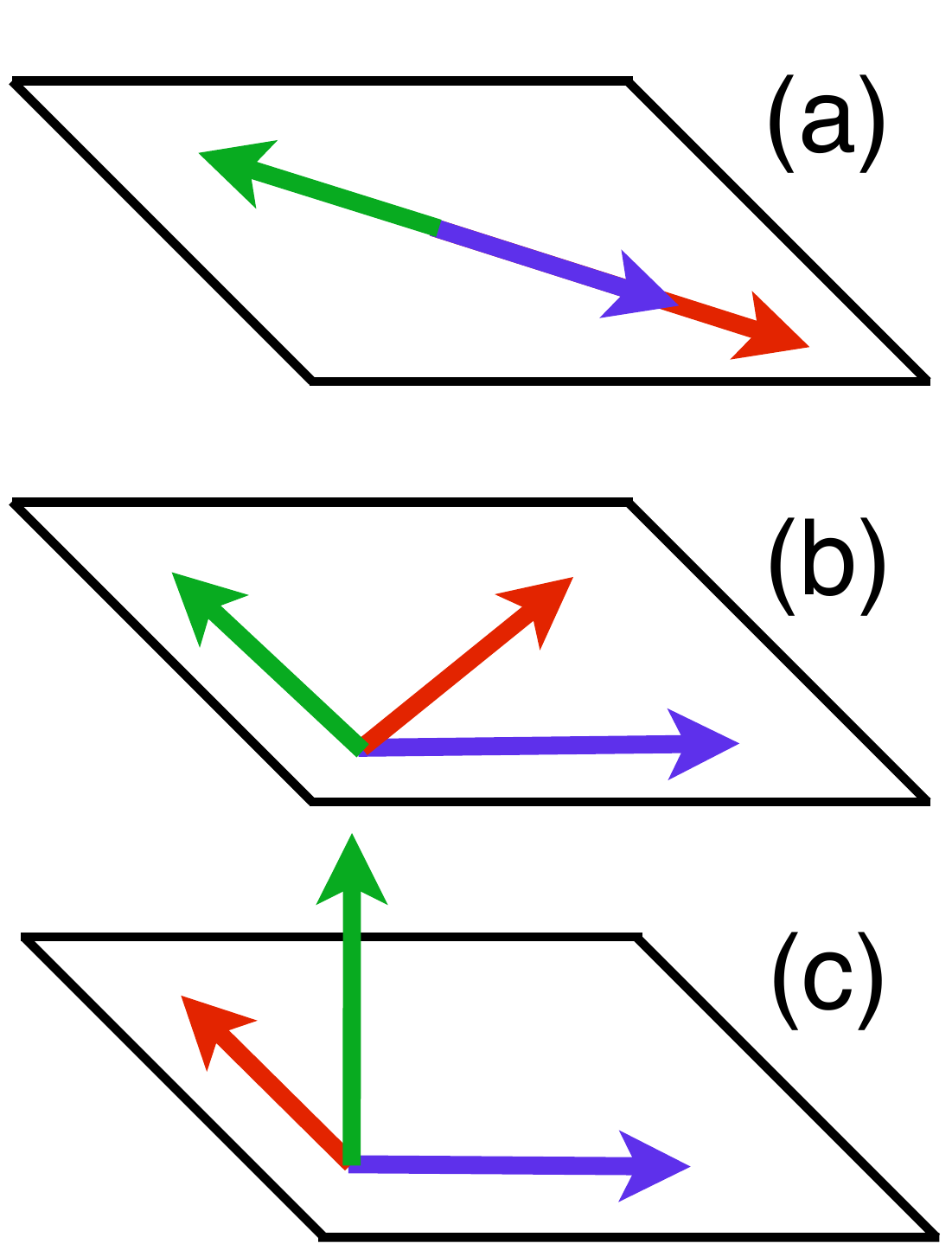}
  \caption{Possible geometrical arrangement of three multi-dimensional vectors $\bar{h}_{x,y,z}$
  (green, blue, and red arrows): (a) collinear arrangement [$l(k)=2$], (b) planar arrangement [$l(k)=1$], 
  (c) linearly independent case [$l(k)=0$].}
  \label{fig:hvecs}
\end{figure}
 
Using a set of $l(k)$'s for a given Hamiltonian one can decide how many qubits can be treated using the 
MF procedure, these will be all qubits with $l(k)=2$. Once all of such qubits have been considered, 
the MF partitioning of $l(k)=1$ qubits begin. For $l(k)=1$, 
the Hamiltonian can be split for any of such qubits in two parts:  $\hat H^{(1)} = \hat h' \hat O'_k$ and 
$\hat H^{(2)} = \hat h'' \hat O''_k + \hat h_e$. 
In both parts the $k^{\rm th}$ qubit can be treated using the MF treatment, which allows one
to continue the consideration for $\hat h', \hat h''$ and $\hat h_e$.
Finally, if only qubits with $l(k)=0$ are left,
then $\hat H$ needs to be partitioned to three Hamiltonians 
$\hat H^{(1)} = \hat h_x \hat x_k$, $\hat H^{(2)} =  \hat h_y \hat y_k$, and
$\hat H^{(3)} =  \hat h_z \hat z_k+ \hat h_e$,    
where at least the $k^{\rm th}$ qubit can be treated using MF.  
After this separation one can apply the reduction chain to each of the three operators. 
Figure~\ref{fig:mfp} illustrates the partitioning for a three qubit case detailed in appendix B.  
  \begin{figure}[h!]
  \centering %
  \includegraphics[width=0.75\columnwidth]{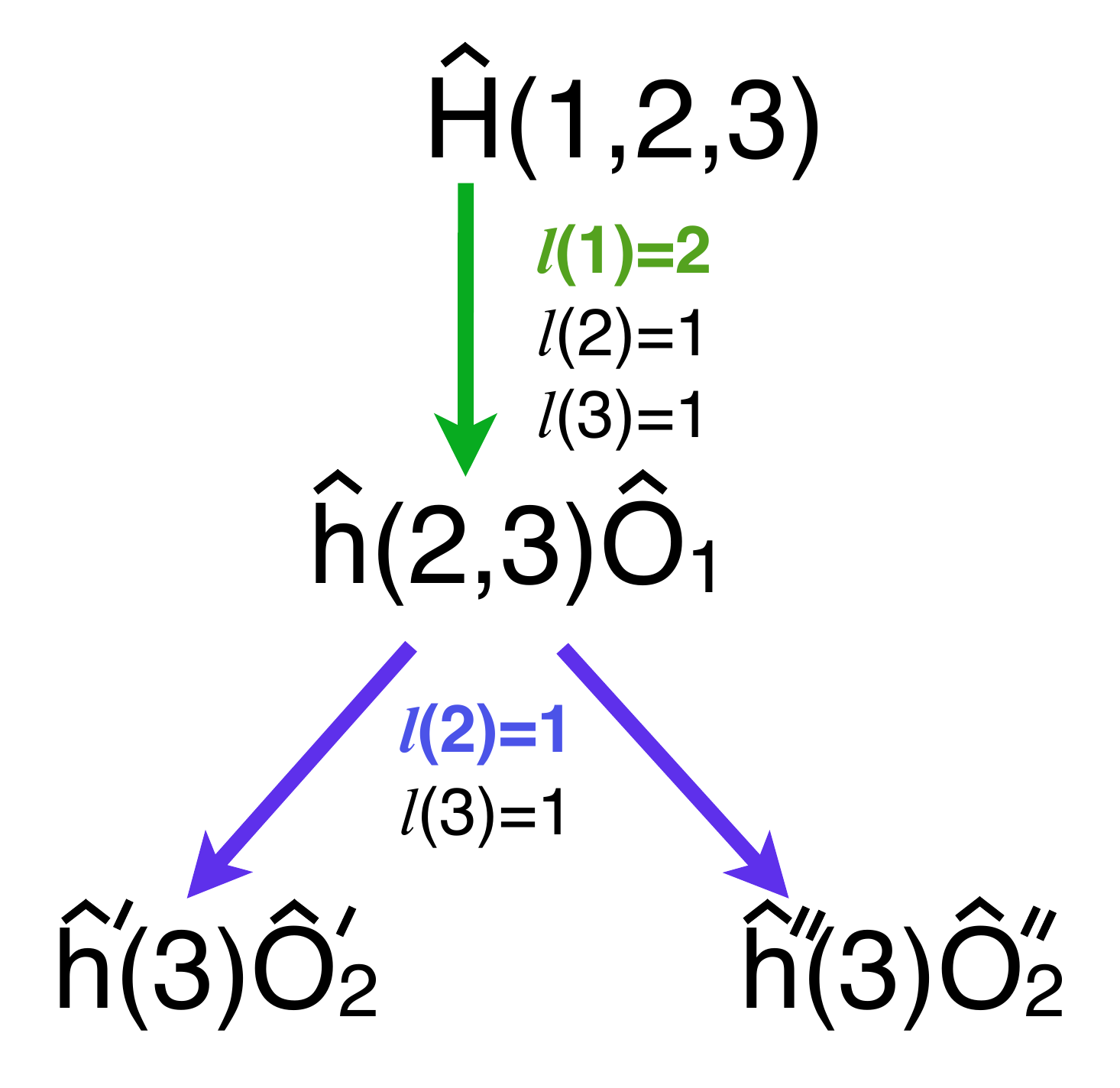}
  \caption{The MF partitioning scheme uses the $l(k)$ function at each step to split a three qubit Hamiltonian (detailed in appendix B)  
  into two fragments. The MF partitioned form is $\hat H(1,2,3) = \hat h'(3)\hat O_2' \hat O_1+\hat h''(3)\hat O_2''\hat O_1$, 
  where all qubits in both fragments can be treated using the MF procedure.}
  \label{fig:mfp}
\end{figure}
In the case when reducing the $k^{\rm th}$ qubit does not produce a Hamiltonian with reducible qubits 
the partitioning needs to be repeated, as in \fig{fig:mfp} when transforming qubit 1 led to $h(2,3)$ where none of the 
qubits can be reduced. 

Our scheme can be considered as an example of greedy algorithm because at every step it 
tries to find locally the most optimal reduction, a qubit with the highest $l(k)$. The reduction 
is only possible if there is linear dependency between complementary vectors $\vec{h}_{x,y,z}$.
The lower the dimensionality of the linear space, where these vectors are located, the more probable 
such linear dependence. Thus, treating qubits with the highest $l(k)$ first is justified 
by the reduction of the space dimensionality along the reductive scheme. In the example of 
\fig{fig:mfp} treatment of qubits 2 and 3 in the beginning would require partitioning of the Hamiltonian to two branches 
for each of them, while leaving the 3$^{\rm rd}$ qubit to the end did not generate any new terms for it. 

It is possible that more than one qubit will have highest $l(k)$. To do more optimal selection in this case, one would 
need to consider maxima of $l(k)$ functions on qubits that enter complementary Hamiltonians $\hat h$
for different reduction candidates. This consideration makes the partitioning computationally costly and was not 
performed in this work.   
  
Applying the partitioning scheme guarantees to result in a sum of MF Hamiltonians that 
can be measured in $N$-qubit one-particle measurements. Since any linear combination of 
QWC terms form a MF Hamiltonian, this partitioning scheme cannot produce
more terms than those used in the regular VQE measuring scheme.   

}

\subsection{Unitary transformations generating mean-fields}
\label{sec:UTMF}

Partitioning the non-MF blocks in the Hamiltonian to obtain more MF terms leads to growth of the 
terms needed to be measured. An alternative treatment of non-MF groups is to search for multi-qubit 
operators that commute with them. Finding such operators may lead to unitary transformations 
that can transform non-MF Hamiltonians into Hamiltonians where qubits shared with 
the commuting operator can be treated using the mean-field procedure. 
{\BC Similar search for multi-qubit operators commuting with the system Hamiltonian was used recently 
by Bravyi and coworkers to reduce the qubit count in in conventional VQE scheme.\cite{Bravyi:2017wb}} 

Let us consider an example where $N$-qubit non-MF Hamiltonian 
$\hat H$ has a two-qubit operator $\hat O^{(2)}(1,2)$ commuting with it (without loss of generality 
we can assume that $\hat O^{(2)}$ acts on the first two qubits). Then, under certain conditions 
detailed in appendix A, $\hat H$ allows for its eigenstates $\Psi$ to be 
written as $\Psi(1,...N) = \Phi(1,2)\psi(3,...N)$, where $\Phi(1,2)$ is an eigenstate of $\hat O^{(2)}$. 
One can always write $\Phi(1,2) = \hat U(1,2) \phi_1(1)\phi_2(2)$, where $\hat U(1,2)$ is an operator 
entangling the product state $\phi_1(1)\phi_2(2)$ into $\Phi(1,2)$. Using this unitary operator, one can 
obtain the Hamiltonian $\hat H_{12} = \hat U(1,2)^{\dagger} \hat H \hat U(1,2)$ that has an eigenstate
$\Psi_{12}(1,...N) = \phi_1(1)\phi_2(2)\psi(3,...N)$ where qubits $1$ and $2$ are unentangled. Therefore,
there should be one-particle operators of qubits $1$ and $2$ that commute with $\hat H_{12}$ and 
its MF-reduced counterpart. Finding these operators and their eigenfunctions $\phi_1(1)$ and $\phi_2(2)$ 
allows us to integrate out qubits $1$ and $2$
\bea
\hat H_{N-2} = \bra{\phi_1\phi_2}H_{12}\ket{\phi_1\phi_2}.
\eea
Search for one- or multi-qubit operators commuting with $\hat H_{N-2}$ can be continued. 
The procedure to find commuting operators with increasing number of qubits requires 
exponentially increasing number of variables parametrizing such 
operators. {\BC Indeed, $k$-qubit operator requires $3^k$ coefficient for all Pauli words
in commutation equations similar to \eq{eq:comm}, also 
the number of different $k$-qubit operators among $N$ qubits is $C_N^k\sim N^k$.}  
Potentially, such operators always exist (e.g., projectors on eigenstates of the Hamiltonian) 
but the amount of resources needed for their search can exceed available. Thus we recommend 
to interchange this search with the partitioning described above if the multi-qubit 
search requires going beyond 2-qubit operators.   

To illustrate the complete scheme involving multi-qubit transformations, 
let us assume that we can continue the reduction chain for $\hat H=\hat H_N$ 
by generating the set of Hamiltonians  $\{\hat H_N,\hat H_{N-2},...,\hat H_{k}\}$ 
using qubit unitary transformations $\{U(1,2), U(3,4,5),...,U(N-k,...N)\}$ and integrating out variables
from $N$ to $k$. To take advantage of this reduction chain in measuring an expectation value of an arbitrary 
wavefunction $\chi(1,...N)$ on $\hat H$, such a measurement should be substituted 
by the following set of conditional measurements:

{\it Step 1:} First two qubits are measured using $\hat H_{12}$ and the unitary transformed function 
$\ket{\hat U(1,2)^{\dagger}\chi}$ because
\bea\notag
\bra{\chi} \hat H \ket{\chi} &=& \bra{\chi} \hat U(1,2) \hat H_{12} \hat U(1,2)^{\dagger} \ket{\chi} \\
&=& \bra{\hat U(1,2)^{\dagger}\chi} \hat H_{12} \ket{\hat U(1,2)^{\dagger}\chi}.
\eea
Depending on the results of these measurements the operator $\hat H_{N-2}$ is formulated and its 
unitary transformation $U(3,4,5)$ is found. $U(3,4,5)$ gives rise to the transformed Hamiltonian 
$\hat H_{35} = \hat U(3,4,5)^{\dagger} \hat H_{N-2} \hat U(3,4,5)$. The wavefunction after measuring 
qubits 1 and 2 is denoted $\ket{\chi_{12}}$. 

{\it Step 2:} Qubits 3-5 are measured on $\hat H_{35}$ sequentially using the transformed wavefunction 
$\hat U(3,4,5)^{\dagger}\ket{\chi_{12}}$. Results of these measurements will define the next 
reduction step and the wavefunction that should be unitarily transformed for the next measurement. 

These steps can be continued until all qubits have been measured. If resources allows for
 finding corresponding multi-qubit unitary transformations, 
 the $\hat H$ Hamiltonian can be measured in $N$ single-qubit measurements.   

\section{Numerical studies and discussion}
\label{sec:numer-stud-disc}

To assess our developments we apply them to the Hamiltonians of the 
\ce{H2} and \ce{LiH} molecules obtained within the STO-3G basis and used to
illustrate performance of quantum computing techniques
previously.\cite{Kandala:2017/nature/242,Hempel:2018/prx/031022,Ryabinkin:2018/qcc}.

\subsection{H$_2$ molecule}
\label{sec:H2_results}

The BK transformed qubit Hamiltonian contains the following terms 
\bea\notag
\hat H_{H_2} &=& C_0 + C_1\hat z_2 + C_2\hat z_3 + C_3\hat z_4 \\ \notag
&+&  C_4\hat z_1 \hat z_3+ C_5\hat z_2 \hat z_4 + C_6\hat z_3 \hat z_4 \\\notag
&+&  C_7\hat z_1 \hat z_2 \hat z_3+ C_8(1+\hat z_1)\hat z_2 \hat z_3 \hat z_4 + C_9\hat z_1\hat z_2 \hat z_4 \\
&+&  C_{10}(1+\hat z_1)\hat y_2 \hat z_3 \hat y_4+ C_{11}(1+\hat z_1)\hat x_2 \hat z_3 \hat x_4. 
\eea
Where some of the $C_i$'s are equal, but it is not going to be important for us 
(the details of generating this Hamiltonian are given in appendix C).
Clearly $\hat H_{H_2}$ contains three groups of QWC terms, first three lines form one group, 
and two last terms fall into two other groups. $\hat H_{H_2}$ is not a MF Hamiltonian, only qubits 1 and 3  
have one-particle operators commuting with the Hamiltonian, while after their reduction the reduced Hamiltonian
does not commute with any one-particle operator
\bea\notag
\hat H_{24} &=& D_0 + D_1\hat z_2 +D_2\hat z_4 + D_3 \hat z_2\hat z_4 \\
&+& D_4 \hat x_2 \hat x_4 + D_5 \hat y_2 \hat y_4,
\eea
where $D_i$'s are constants. 
Partitioning of $\hat H_{24}$ to three terms using qubit 2 or 4 
would not be more efficient than partitioning $\hat H_{H_2}$ in 3 groups of QWC terms from the beginning. 
{\BC
However, there is the two-particle operator $\hat z_2 \hat z_4$ that commutes with $\hat H_{24}$, 
and it can be used to devise a unitary transformation bringing $\hat H_{24}$ to the MF form. 
Note, that even though $\hat z_2 \hat z_4$ has a spectral degeneracy, this degeneracy will not create 
problematic entanglement discussed in appendix A,
because there are no other qubits besides 2 and 4 in $\hat H_{24}$.
The sought unitary transformation is $U(2,4)=\exp[-i (3\pi/2) \hat z_2 \hat x_4]$, 
and the transformed MF Hamiltonian is} 
\bea\notag
U(2,4)^{\dagger}\hat H_{24} U(2,4) &=& E_0 + E_1 \hat z_2 + E_2 \hat y_2 + E_3 \hat y_4 \\ \label{eq:MFH}
&+& E_4 \hat y_2 \hat y_4 + E_5 \hat z_2 \hat y_4,   
\eea
where $E_i$'s are some constants and the first one-particle commuting operator is $\hat O_1(4) = \hat y_4$. 
After integrating out $\hat O_1$'s eigenfunction, $\hat O_2(2)$ is a linear combination of $\hat z_2$ and 
$\hat y_2$.

To illustrate superiority of the scheme with use of $U(2,4)$ and measurements 
of the MF Hamiltonian over the regular approach with splitting $\hat H_{H_2}$ to 
three groups of QWC operators, Table~\ref{tab:var} presents variances 
for the Hamiltonian expectation value for two wavefunctions, the exact eigenfunction ($\Psi_{\rm QCC}$) 
of and the mean-field approximation ($\Psi_{\rm QMF}$) to the ground state 
of the H$_2$ problem at $R$(H-H)$=1.5$ \AA.\cite{Ryabinkin:2018/qcc} 
The exact solution measured in the new scheme (MF-partitioning 2p) gives only one value 
with zero variance, while the regular schemes gives three distributions for each non-commuting term.

\begin{table}  
  \centering
  \caption{Estimates of total variances ($Var$) for the H$_2$ and LiH molecules with different 
  partitioning approaches and wavefunctions ($\Psi_{\rm QCC}$ from the 
  qubit coupled cluster method,\cite{Ryabinkin:2018/qcc} 
  and $\Psi_{\rm QMF}$ from the qubit mean-field approach\cite{Ryabinkin:2018gx} ). 
The number of terms corresponds to the number of separately measured $N$-qubit terms. 
  For all partitionings, covariances have not been included in the $Var$ estimates, which simulates 
  practical estimation of the total variance.}
  \label{tab:var}
    \begin{tabular}{@{}cccc@{}}
    Approach & Number of terms & $Var(\Psi_{\rm QCC})$ & $Var(\Psi_{\rm QMF})$ \\
    \hline
    \multicolumn{4}{c}{\it H$_2$}\\
    QWC-partitioning & 3 & 0.044  & 0.026  \\
    MF-partitioning 2p & 1 & 0 & 0.053  \\
    $\langle\hat H_{\rm H_2}^2\rangle-\langle\hat H_{\rm H_2}\rangle^2$ & 1 & 0 & 0.053  \\
        \multicolumn{4}{c}{\it LiH}\\
    QWC-partitioning & 25 & 0.043 & 0.037 \\
    MF-partitioning 1p & 13 & 0.029 & 0.036 \\
    MF-partitioning 2p & 5 & 0.030 & 0.038 \\
        $\langle\hat H_{\rm LiH}^2\rangle-\langle\hat H_{\rm LiH}\rangle^2$ & 1 & $5.6\times10^{-4}$ & 0.027 \\
     \end{tabular}
\end{table}
{\BC In the approximate wavefunction case, the true variance obtained from the Hamiltonian
 is larger than that of the conventional approach. This is a consequence of ignoring covariances in the conventional 
 approach. The MF partitioning 2p variance is equal to the exact one, since it is obtained from measuring 
 a single term (the MF Hamiltonian in \eq{eq:MFH}) and thus does not neglected any covariances.}

\subsection{LiH molecule}
\label{sec:LiH_results}
We will consider the LiH molecule at $R{\rm (Li-H)}=3.2$ \AA, it 
has a 6-qubit Hamiltonian containing 118 Pauli words 
(see appendix C for details). This qubit Hamiltonian has 3$^{rd}$ and 6$^{th}$
stationary qubits, which allows one to replace the corresponding
$\hat z$ operators by their eigenvalues, $\pm 1$, thus defining the
different ``sectors'' of the original Hamiltonian. Each of these
sectors is characterized by its own 4-qubit effective Hamiltonian. The
ground state lies in the $z_{3} = -1$, $z_{6} = 1$ sector; 
the corresponding 4-qubit effective Hamiltonian ($\hat H_{\rm LiH}$) has
100 Pauli terms.  
Integrating out 3$^{rd}$ and 6$^{th}$ qubits can be done in the MF 
framework. The MF treatment of $\hat H_{\rm LiH}$ is not possible without its partitioning. 

Before discussing partitioning of $\hat H_{\rm LiH}$ we would like to note that 
there are two 2-qubit operators commuting with 
 $\hat H^{(4)}$ (we reenumerate qubits after the reduction from 6 to 4 qubits in the Hamiltonian)
\bea
\hat O_1^{(2)} &=&  -\hat z_1 + \hat z_2 - \hat z_1\hat z_2 \\
\hat O_2^{(2)} &=&  -\hat z_3 + \hat z_4 + \hat z_3\hat z_4.
\eea
Unfortunately, both operators have degenerate spectra with a single non-degenerate eigenstate
and three degenerate states. Moreover, these degeneracies do not satisfy the factorability condition 
introduced in appendix A and thus proves impossible to find 2-qubit unitary transformation
that would factorize qubits 1 and 2 or 3 and 4.  

Table~\ref{tab:var} summarizes results of partitioning for $\hat H_{\rm LiH}$ and variances calculated for different
wavefunctions and partitioning schemes. The partitioning involving only one-qubit transformations 
(MF-partitioning 1p) reduces the number of QWC terms by half. Involving the two-qubit transformations 
at the step before the last one in the MF partitioning reduces the number of terms to only 5 (MF-partitioning 2p),
which is a fivefold reduction compare to the conventional QWC form.  
{\BC Alternative pathways in the MP partitioning scheme related to different choices of partitioned qubits with the same 
value of $l(k)$ generated not more than 15 and 9 terms for MF partitioning 1p and 2p, respectively.} 
As previously, the qubit mean-field ($\Psi_{\rm QMF}$) 
and qubit coupled cluster ($\Psi_{\rm QCC}$) wavefunctions are considered, with only difference that 
$\Psi_{\rm QCC}$ is a very accurate but not exact ground state wavefunction for LiH 
(thus there is a small but non-zero variance of the $\hat H_{\rm LiH}$ on $\Psi_{\rm QCC}$). 
Details on generation of these functions can be found in Ref.~\citenum{Ryabinkin:2018/qcc}. 
Variances across different partitionings do not differ appreciably and the main advantage of the 
MF-partitioning schemes is in the reduction of the number of terms that need to be measured. 
 
\section{Conclusions}
\label{sec:conclusions}

We have introduced and studied a new method for partitioning of the qubit Hamiltonian
in the VQE approach to the electronic structure problem. The main idea of our approach is 
to find Hamiltonian fragments that have eigenstates consisting of single products of one- and 
two-qubit wavefunctions. The most general criterion for identifying such Hamiltonian fragments 
was derived for the first time. Once such fragments are found the total 
wavefunction of the system can be measured on a fragment Hamiltonian in a single pass 
of $N$ single-qubit measurements intertwined with one- and two-qubit rotations that are defined 
on-the-fly from results of previous qubit measurements. The main gain from such a 
reformulation is a decrease of separately measured Hamiltonian fragments.    
Indeed, illustrations on simple molecular systems (H$_2$ and LiH) shown three- and five-fold reductions 
of the number of terms that are needed to be measured with respect to the conventional scheme. 

In the process of deriving our partitioning procedure, we discovered criteria for eigenstate
factorability for an arbitrary Hamiltonian acting on $N$ distinguishable particles. 
Our criteria involve search for few-body operators commuting with the Hamiltonian of interest. 
Even though the criteria for factorability are exact, realistic molecular Hamiltonians do not satisfy them in 
general. Therefore, we needed to introduce  
a heuristic partitioning procedure (greedy algorithm) that splits the system Hamiltonian to fragments that have 
factorable eigenstates. Even though the procedure does not guarantee the absolutely 
optimal partitioning to the smallest number of terms, it does not produce more terms 
than the number of qubit-wise commuting sub-sets.   

Interestingly, when one is restricted with single-qubit measurements, commutation property 
of two multi-qubit operators $\hat A$ and $\hat B$ has nothing to do with ability to measure 
them together (see Table~\ref{tab:com}). This seeming contradiction with laws of quantum 
mechanics arise purely from a hardware restriction that one can measure a single qubit at a time. 
On the other hand, qubit-wise commutativity is still a sufficient but not necessary 
condition for single-qubit measurability. 
Removing the single-qubit measurement restriction in the near future will 
not make our scheme obsolete but rather would allow
to skip the single-particle level. For example, if two-qubit measurements will be available, 
one can look for two-qubit  operators commuting with the Hamiltonian and integrate out pairs of 
qubits to define next measurable two-qubit operators.   
\begin{table}  
  \centering
  \caption{Commutativity of two operators  and their simultaneous single-qubit measurability (SQM).}
  \label{tab:com}
    \begin{tabular}{@{}cccc@{}}
    $\hat A$ & $\hat B$ & $[\hat A,\hat B]$ & SQM of $(\hat A + \hat B)$ \\
    \hline
    $\hat z_1\hat z_2$ & $\hat z_2\hat z_3$ & 0 & Yes \\
    $\hat z_1\hat z_2$ & $\hat x_1\hat x_2$ & 0 & No \\
        $\hat z_1\hat z_3$ & $\hat x_1\hat z_2$ & $\ne 0$ & Yes \\
            $\hat z_1\hat z_2$ & $\hat x_1\hat y_2$ & $\ne 0$ & No 
    \end{tabular}
\end{table}

The current approach can address difficulties arising in exploration of the excited state via 
minimization of variance
\bea\label{eq:varE}
E(\omega) = \min_{\theta} \bra{\Psi(\theta)} (\hat H - \omega)^2\ket{\Psi(\theta)}. 
\eea
One of the largest practical difficulties is in an increasing number of terms that are required to be measured
in \eq{eq:varE}. Combining some of these terms using the current methodology can reduce the number of 
needed measurements.

A similar problem with a growing number of terms arise if one would like to obtain the true quantum uncertainty 
of the measurements for a partitioned Hamiltonian, it requires measuring all covariances between all parts. 
Ignoring covariances by assuming measurement independence can lead to incorrect 
estimation of the true uncertainty, both under- and over-estimation are possible.     
 
 {\BC
From the hardware standpoint, the new scheme requires modification of the single-qubit 
measurement protocol, where measurement results for some qubits will define unitary rotations of 
other qubits before their measurement, so-called {\it feedforward} measurement. This type of measurement
has already been implemented in quantum computers based on 
superconducting\cite{Vijay:2012bv} and 
photonics\cite{Prevedel:2007ca,Moqanaki:2015iw,Reimer:2018cv} 
qubit architectures in the context of measurement-based quantum computing.\cite{Nielsen:2006ve,Mantri:2017el} 
}
Thus we hope that the new method will become the method of choice for quantum
chemistry on a quantum computer in near future. 


\section*{Acknowledgement}
A.F.I. is grateful to D. Segal, P. Brumer, R. Kapral, J. Schofield, J. R. McClean,
D. F. James, and I. Dhand for useful discussions.  
A.F.I. acknowledges financial support from the Natural Sciences and
Engineering Research Council of Canada. 

%

\appendix
\section*{Appendix A: Factorization conditions for the Hamiltonian eigenstates}
\label{sec:proof}
Here we prove that the given in the main text condition for 
a $N$-qubit Hamiltonian to be in the MF class is actually a necessary and sufficient condition, 
and hence is a criterion. We will split the proof in two parts: 
1) If the Hamiltonian has $N$ one-particle operators satisfying the reduction chain, its eigenfunctions 
can be written as products (sufficiency); 
2) If all the Hamiltonian eigenfunctions are in a product form then it will have $N$
commuting one-particle operators defined by the reduction scheme (necessity). 

{\it 1) Proof of sufficiency:} If there exist $N$ one-particle operators commuting with a set of reduced 
Hamiltonians it is straightforward to check that a product of eigenstates of these operators is an eigenstate 
of the Hamiltonian. Note that any nontrivial one-qubit operator has a non-degenerate spectrum, therefore, 
there is no degree of freedom related to rotation within a degenerate subspace. 
The choice of the first eigenstate of the 
first operator ($\hat O_1$) can define the form of next one-particle operators and their eigenstates.    

{\it 2) Proof of necessity:} For the $N$-particle eigenstate $\Psi(1,...N)$ to have a product form it is necessary 
for the Hamiltonian to have eigenstates of the $\phi_1(1)\Phi(2,...N)$ form, where $\phi_1(1)$ and 
$\Phi(2,...N)$ are some arbitrary functions from Hilbert spaces of qubit 1 and $N-1$ qubits. 
The later form is an eigenstate of an operator of the form $\hat O_1\otimes I_{N-1}$, where $I_{N-1}$ is an 
identity operator and $\hat O_1$ is an operator for which $\phi_1(1)$ is an eigenfunction. 
Then, if the Hamiltonian and $\hat O_1\otimes I_{N-1}$ share the eigenstates they must commute. This 
commutation is equivalent to $[\hat H, \hat O_1]=0$. The same logic can be applied to $\Phi(2,...N)$ because 
the next necessary condition for the total eigenfunction of the Hamiltonian to be in a product form is that 
$\Phi(2,...N) = \phi_2(2)\tilde{\Phi}(3,...N)$, this gives rise to another commuting operator $\hat O_2$ whose 
eigenfunction is $\phi_2$. It is important to note though that $\hat O_2$ does not need to commute with 
$\hat H$ but only with its reduced version $H_{N-1} = \bra{\phi_1}\hat H\ket{\phi_1}$. This chain can be 
continued until we reach the end of the variable list.

\paragraph*{Many-particle commuting operator extension:} Similarly if we can find $M$-particle operator $\hat O$
commuting with $\hat H$ then, because of the theorem about commuting operators, there is a common 
set of eigenfunctions. With multi-qubit operators one needs to be careful because they can have degenerate 
spectrum. In the case of non-degenerate spectrum of $\hat O$ the common 
eigenstates have the factorized form $\Psi(1,...N) = \Phi(1,...M)\chi(M+1,...N)$, which serves a solid ground 
for the discussion in the main text. In the degenerate case, the most general form of a common eigenstate
is $\Psi(1,...N) = \sum_{I} C_I\Phi_I(1,...M)\chi_I(M+1,...N)$, where 
$\hat O \Phi_I(1,...M) = \lambda \Phi_I(1,...M),~I=1,...k$. In this case, the important question becomes whether 
the Hamiltonian allows for the eigenstates to be single product states,
\bea\notag
\Psi(1,...N) = \left[\sum_{I} C_I\Phi_I(1,...M)\right]\chi(M+1,...N),
\eea
or not? To answer this question one needs to construct a reduced matrix operator within the degenerate 
subspace $\{\Phi_I(1,...M)\}$
\bea
\hat H_{IJ}^{(N-M)} = \bra{\Phi_I} \hat H \ket{\Phi_J},
\eea
where integration is done over the first $M$ variables.
If there exists $\Phi_I$ for which $\hat H_{IJ}^{(N-M)}=0$ where $J\ne I $ then 
$\Psi(1,...N) = \Phi_I(1,...M)\chi(M+1,...N)$ will be an eigenfunction of the Hamiltonian. 
For all $\Phi_I(1,...M)$ eigenfunctions to form product states, all off-diagonal 
elements of  $\hat H_{IJ}^{(N-M)}$ must be zero. 
There is one more possibility for the factorized eigenstates, 
if the reduced matrix operator has the particular form 
\bea
\hat H_{IJ}^{(N-M)} = h_{IJ} \hat H^{(N-M)},
\eea
where $h_{IJ}$ are elements of a constant matrix and $\hat H^{(N-M)}$ is a single reduced operator
acting on $N-M$ variables. 
Note that for doing this analysis one needs to be able to obtain only eigenstates of $\hat O$. This is 
presumably easier procedure since $M<N$. 
 
Thus, in the degenerate case, having a product form is not guaranteed and therefore, one may 
be able to obtain the unitary transformation unentangling qubits only in the described two cases. 
Yet, finding the commuting operator 
$\hat O$ is a necessary condition for existence of an unentangling unitary transformation.    

{\BC

\section*{Appendix B: Illustration of the mean-field partitioning procedure}

    To illustrate the MF partitioning procedure on a nontrivial example let us consider the model Hamiltonian whose partitioning 
    gives rise to the scheme on \fig{fig:mfp}
    \bea
    \hat H &=& 3\hat x_1\hat x_2\hat x_3 + \hat x_1\hat x_2\hat y_3 + 5\hat x_1\hat x_2\hat z_3 + 5\hat x_1\hat y_2\hat x_3 + 7\hat x_1\hat y_2\hat z_3 \notag \\
    && + 3\hat x_1\hat z_2\hat x_3 + \hat x_1\hat z_2\hat y_3 + 5\hat x_1\hat z_2\hat z_3 + 6\hat y_1\hat x_2\hat x_3 + 2\hat y_1\hat x_2\hat y_3  \notag \\ 
    &&+ 10\hat y_1\hat x_2\hat z_3 + 10\hat y_1\hat y_2\hat x_3 + 14\hat y_1\hat y_2\hat z_3 + 6\hat y_1\hat z_2\hat x_3 + 2\hat y_1\hat z_2\hat y_3 \notag \\ &&
    + 10\hat y_1\hat z_2\hat z_3 + 3\hat z_1\hat x_2\hat x_3 + \hat z_1\hat x_2\hat y_3 + 5\hat z_1\hat x_2\hat z_3 + 5\hat z_1\hat y_2\hat x_3\notag \\ && + 7\hat z_1\hat y_2\hat z_3 + 3\hat z_1\hat z_2\hat x_3 + \hat z_1\hat z_2\hat y_3 + 5\hat z_1\hat z_2\hat z_3
    \eea
    To assess whether the partitioning of $\hat H$ is possible based on qubit $k=1$ we rewrite the Hamiltonian as  
    \bea
     \hat H = \hat x_1\hat h_x + \hat y_1\hat h_y + \hat z_1\hat h_z,
    \eea where 
    \bea
    \hat h_x &=& 3\hat x_2\hat x_3 + \hat x_2\hat y_3 + 5\hat x_2\hat z_3 + 5\hat y_2\hat x_3 + 7\hat y_2\hat z_3 \notag \\ &&+ 3\hat z_2\hat x_3 + \hat z_2\hat y_3 + 5\hat z_2\hat z_3\\
    \hat h_y &=& 6\hat x_2\hat x_3 + 2\hat x_2\hat y_3 + 10\hat x_2\hat z_3 + 10\hat y_2\hat x_3 + 14\hat y_2\hat z_3 \notag \\ && + 6\hat z_2\hat x_3 + 2\hat z_2\hat y_3 + 10\hat z_2\hat z_3\\
    \hat h_z &=& 3\hat x_2\hat x_3 + \hat x_2\hat y_3 + 5\hat x_2\hat z_3 + 5\hat y_2\hat x_3 + 7\hat y_2\hat z_3 \notag \\ &&+ 3\hat z_2\hat x_3 + \hat z_2\hat y_3 + 5\hat z_2\hat z_3
    \eea Each $\hat h_{x,y,z}$ is transformed into a vector. For example \begin{align}
        \vec{h}_x = \begin{bmatrix} 
          3 & 1 & 5 & 5 & 7 & 3 & 1 & 5
        \end{bmatrix}^T \label{eq:term_basis}
    \end{align} in the basis 
    $\{\hat x_2\hat x_3, \hat x_2\hat y_3,\hat x_2\hat z_3, \hat y_2\hat x_3,\hat y_2\hat z_3,\hat z_2\hat x_3,\hat z_2\hat y_3,\hat z_2\hat z_3\}$. $S_1$ is obtained as $A_1^\dag A_1$, where  $A_1= [\vec{h}_x ~ \vec{h}_{y} ~ \vec{h}_z]$. 
    Diagonalizing of $S_1$ gives one non-zero eigenvalue $d$ and a corresponding eigenvector $\vec{v}$. 
    The dimensionality of the $S_1$ kernel is 2, $l(1)=2$, and it implies collinearity of $\vec{h}_{x,y,z}$ (\fig{fig:hvecs}a). 
    Performing similar analysis for $S_2$ and $S_3$, one can find $l(2)=l(3)=1$ (see \fig{fig:mfp}). 
    Therefore, we rewrite the Hamiltonian as $\hat H = \hat h(2,3)\hat O_1$, where 
 \bea
    \hat O_1 &=& 0.408248\hat x_1 + 0.816497\hat y_1 + 0.408248\hat z_1\\
    \hat h(2,3) &=& 7.34847\hat x_2\hat x_3 + 2.44949\hat x_2\hat y_3 \notag \\ &&+ 12.2474\hat x_2\hat z_3 + 12.2474\hat y_2\hat x_3 \notag \\ &&+ 17.1464\hat y_2\hat z_3 + 7.34847\hat z_2\hat x_3 \notag \\ &&
    + 2.44949\hat z_2\hat y_3 + 12.2474\hat z_2\hat z_3
    \eea $\hat O_1$ and $\hat h(2,3)$ were obtained through linear combination of 
    $\{\hat x_1,\hat y_1,\hat z_1\}$ and $\{\hat h_x, \hat h_y, \hat h_z\}$ with coefficients from the eigenvector $\vec{v}$. 

    As the next step, we consider $\hat h(2, 3)$, it can be partitioned based on either qubit $k = 2$ or $k=3$. 
    Both qubits have the same values of $l(k)=1$ and $\vec{h}_{x,y,z}$ are in a single plane (\fig{fig:hvecs}b). 
    Here, we choose arbitrarily $k=2$, diagonalizing $S_2$ leads to two non-zero eigenvalues $(d_1,d_2)$ 
    and corresponding eigenvectors $(\vec{v}_1,\vec{v}_2)$. Following the procedure, $\hat h(2,3)$ decomposes to 
      \be
      \hat h(2,3) = \hat h'(3)\hat O_2' + \hat h''(3)\hat O_2'',
      \ee where \bea
      \hat O_2' &=& 0.507019\hat x_2 - 0.697039\hat y_2 + 0.507019\hat z_2 \\
      \hat h'(3) &=& -1.08532\hat x_3 + 2.48388\hat y_3 + 0.467647\hat z_3 \\
      \hat O_2'' &=& 0.492881\hat x_2 + 0.717033\hat y_2 + 0.492881\hat z_2 \\
      \hat h''(3) &=& 16.0257\hat x_3 + 2.41461\hat y_3 + 24.3676\hat z_3.
      \eea
 The single-qubit operators $\{\hat O_2', \hat O_2''\}$ and their complements $\{\hat h', \hat h''\}$ were obtained 
 taking linear combinations of $\{\hat x_2, \hat y_2, \hat z_2\}$ and $\{\hat h_x, \hat h_y, \hat h_z\}$ 
 with coefficients from the eigenvectors $(\vec{v}_1,\vec{v}_2)$, respectively. 
 
 The complexity of a single step of the MF partitioning procedure is polynomial with the number of qubits.
 In each step we need to evaluate the $l(k)$ function for each of present qubits. Evaluation of the 
 $l(k)$ function requires building the corresponding overlap matrix $S_k$, which involves inner products 
 between columns of $A_k$ matrices. Since the length of $A_k$ columns ($\bar{h}_{x,y,z}$) 
 scales as $N^4$ at most (this is the scaling of the total number of terms in the Hamiltonian), 
 the construction of $S_k$ scales as $N^4$ as well. Thus funding $l(k)$ functions for all qubits 
 in general has $O(N^5)$ scaling.   

}

\section*{Appendix C: Hamiltonian details}
\paragraph*{H$_2$ molecule:}
One- and two-electron integrals in the canonical \gls{RHF} molecular
orbitals basis for $R$(H-H)=1.5 \AA, 
were used in the \gls{BK} transformation to produce the corresponding qubit Hamiltonian.
Spin-orbitals were alternating in the order $\alpha$, $\beta$,
$\alpha$, .... The explicit expression for the BK qubit Hamiltonian 
is given in Supplementary Information.

\paragraph*{LiH molecule:}
Qubit Hamiltonian for $R$(Li-H)=3.2\AA~ distance 
was generated using the parity fermion-to-qubit transformation\cite{Nielsen:2005/scholar_text}.
Spin-orbitals were arranged as ``first all alpha then all beta'' in
the fermionic form; since there are 3 active molecular orbitals in the
problem, this leads to 6-qubit Hamiltonian. Further details on the Hamiltonian 
are given in Supplementary Information.


\end{document}